\documentclass[12pt,preprint]{aastex}
\usepackage{amsmath}
\newcommand{\kms}          {\mbox{${\rm km~s^{-1}}$}}

\newcommand{\e}            {\mbox{$^{-1}$}}
\newcommand{\ee}           {\mbox{$^{-2}$}}
\newcommand{\eee}          {\mbox{$^{-3}$}}
\def\cm2{\mbox{${\rm cm^{-2}}$}}
\def\h2{\mbox{${\rm H}_2$}}
\def\nh2{\mbox{$n_{\rm H_2}$}}
\def\Nh2{\mbox{$N_{{\rm H}_2}$}}
\def\Mh2{\mbox{$M_{{\rm H}_2}$}}

\def\farcs{\hbox{$.\!\!^{''}$}}

\def\micron{\hbox{$\mu$m}}
\def\simgt{\lower.5ex\hbox{$\; \buildrel > \over \sim \;$}}
\def\simlt{\lower.5ex\hbox{$\; \buildrel < \over \sim \;$}}

\def\t1c{\mbox{$\theta^1$\,Ori\,C}}
\def\aj{AJ}
\def\apj{ApJ}
\def\apjl{ApJ}
\def\13co{$^{13}$CO}
\def\c18o{C$^{18}$O}
\def\H2{H$_2$}

\def\startfigcap{\vspace*{2.0\baselineskip}\bgroup\leftskip 0.45in\rightskip 0.45in\small}
\def\endfigcap{\par\egroup\vspace*{2.0\baselineskip}}

\def\startfigcapside{\vspace*{2.0\baselineskip}\bgroup\leftskip 4.0in\rightskip 0.4in\small}
\def\endfigcapside{\par\egroup\vspace*{2.0\baselineskip}}

\def\h2co{H$_2$CO}

\begin{document}

\title{Diverse protostellar evolutionary states in the young cluster AFGL961}
\author{Jonathan P. Williams$^1$, Rita K. Mann$^1$,
Christopher N. Beaumont$^1$, Jonathan J. Swift$^1$,
Joseph D. Adams$^2$, Joe Hora$^3$, Marc Kassis$^4$,
Elizabeth A. Lada$^5$, Carlos G.  Rom{\'a}n-Z{\'u}{\~n}iga$^6$}
\affil{$^1$ Institute for Astronomy, University of Hawaii, 2680 Woodlawn Drive, Honolulu, HI 96822; jpw@ifa.hawaii.edu}
\affil{$^2$ Cornell University, Department of Radiophysics Space Research, Ithaca NY 14853}
\affil{$^3$ Harvard-Smithsonian, CfA, 60 Garden St., MS 65, Cambridge, MA 02138}
\affil{$^4$ Keck Observatory 65-1120 Mamalahoa Hwy, Kamuela, HI 96743}
\affil{$^5$ Department of Astronomy, University of Florida, Gainesville, FL 32611}
\affil{$^6$ Centro Astron\'omico Hispano Alem\'an, Camino Bajo de Hu\'etor 50, Granada 18008, Spain}
\shorttitle{Diverse evolutionary states in AFGL961}
\shortauthors{Williams et al.}

\begin{abstract}
We present arcsecond resolution mid-infrared and millimeter observations of the
center of the young stellar cluster AFGL961 in the Rosette molecular cloud.
Within 0.2\,pc of each other, we find an early B star embedded in a dense core,
a neighboring star of similar luminosity with no millimeter counterpart,
a protostar that has cleared out a cavity in the circumcluster envelope,
and two massive, dense cores with no infrared counterparts.
An outflow emanates from one of these cores,
indicating a deeply embedded protostar, but the other is
starless, bound, and appears to be collapsing.
The diversity of states implies either that protostellar evolution is
faster in clusters than in isolation or that clusters form via
quasi-static rather than dynamic collapse. The existence of a
pre-stellar core at the cluster center shows that that some star formation
continues after and in close proximity to massive, ionizing stars.
\end{abstract}

\section{Introduction}
The dominant mode of star formation is in groups.
In this sense, understanding cluster formation is
a prerequisite for understanding the origin of most stars,
including all massive stars and, in all likelihood, our Sun.
Some major unanswered questions include the role of global and local
processes, the formation time-scale, and whether high mass
stars shut down further cluster growth.

Long wavelength, mid-infrared to millimeter, observations
are required to image the youngest, most deeply embedded
protostars and their nascent cores. The study of dense
stellar groups is hampered by the low resolution at these
wavelengths and, consequently, less is known about their
origins and evolution than closer, isolated star forming systems.
Technological developments including large format
mid-infrared arrays and sensitive millimeter interferometers
provide a new view of the most embedded regions of young clusters.
We report here observations of young protostars and
molecular cores in the AFGL961 cluster and the
constraints that these place on models of cluster formation.

AFGL961 owes its name to an Air Force infrared sky survey
carried out on a rocket-borne telescope in the early 1970s.
The resolution of these data was low and \cite{Cohen73}
was the first to give a precise, ground-based, position.
Due to its high luminosity and location in the nearby
Rosette molecular cloud, AFGL961 has been the subject of many
subsequent studies.

\cite{Bally83} found a weak point-like 5\,GHz source with the
VLA and interpreted it as a compact HII region around a B3 star.
Based on the definition of \cite{Zinnecker07},
AFGL961 can therefore be considered as a (borderline)
massive star forming region.
Near-infrared imaging by \cite{Lenzen84} revealed a double
source with the easternmost, redder component coinciding
with the VLA source. They also found a third object
about 30'' further west near the most prominent optical
emission, a fan-shaped reflection nebula.
\cite{Castelaz85} confirmed the double source and showed that
both members are pre-main-sequence with the luminosity of B stars.
\cite{Aspin98} presented a wide field image of shocked H$_2$
emission showing numerous stellar sources and bow shock structures.
\cite{Roman08} mapped the entire Rosette molecular cloud
in the near-infrared and identified 10 embedded clusters of
which AFGL961 is the brightest and the most heavily obscured by
nebulosity. This obscuration, however, prevented a detailed
study of the fainter and most embedded sources.

AFGL961 lies in a massive clump along a broad ridge of molecular
emission extending away from the Rosette nebula \citep{Blitz80}.
\cite{Lada82} mapped a bipolar outflow from the cluster,
most recently imaged in CO 3--2 by \cite{Dent09} who show a
high velocity, collimated flow extending over $5'$.

Distance estimates to the Rosette cloud range from $1.4-1.7$\,kpc
\citep{Ogura81,Perez87,Hensberge00}
and we adopt 1.6\,kpc here for consistency with most previous work.
AFGL961 has an infrared luminosity of $11400\,L_\odot$ \citep{Cox90}
and is, by far, the brightest embedded cluster in the cloud.

\begin{figure}[t]
\vskip -1in
\includegraphics[width=5in,angle=90]{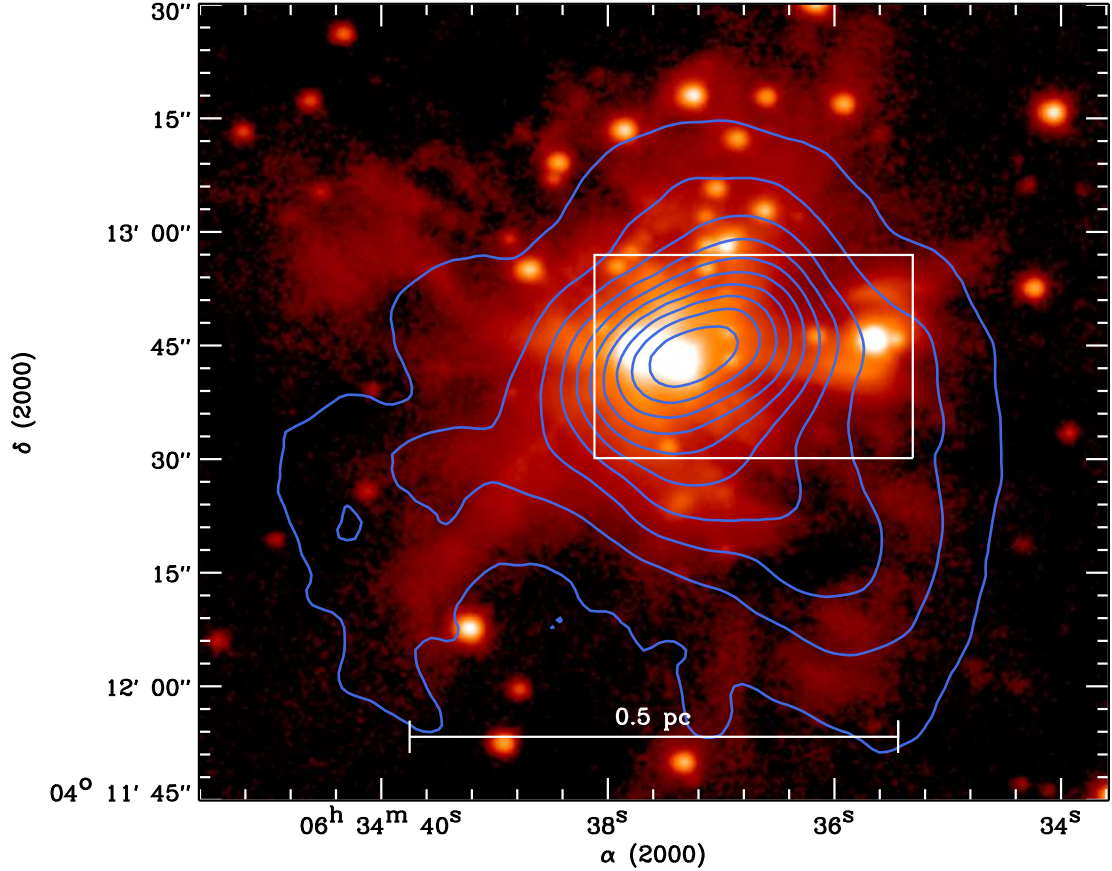}
\caption{Large scale view of the AFGL961 cluster. The background
is the K-band image from \cite{Roman08} on a log scale showing
the embedded stars and associated nebulosity.
The contours of the SCUBA 850\,\micron\ emission from
the cold, dusty cluster envelope show surface densities at
$\Sigma=0.025\times(1,2,3...)$\,g\,cm\ee.
The rectangle outlines the region shown in the MIRSI images in Figure~2.}
\label{fig.flamingos}
\end{figure}

A large-scale view of AFGL961 is shown in Figure~\ref{fig.flamingos}.
The image is from the observations of \cite{Roman08} and
shows the embedded protostars and associated nebulosity at 2.2\,\micron.
The contours show the emission at 850\,\micron\ from the dusty cluster
envelope and were produced from archival SCUBA data.
The flux per beam at this wavelength is a direct measure of column
density where we have assumed a uniform temperature $T=20$\,K and
dust opacity $\kappa_\nu=0.1(\nu/1200\,{\rm GHz})$\,cm$^2$\,g$^{-1}$
\citep{Hildebrand83}.
The contours are in units of g\,cm\ee\ to best compare
to theoretical calculations and numerical simulations (see \S4).
The total flux of the map is 10\,Jy which converts to a
mass $M=90\,M_\odot$.
Three bright stars dominate the luminosity of the cluster
at this wavelength and through the mid-infrared \citep{Poulton08}.
The dust peak is slightly elongated and offset from the
central double source and the western source at the center
of the optical fan-shaped nebulosity lies near the edge of the envelope.

We have gathered data on AFGL961 over a period of several years
with the Infrared Telescope Facility\footnotemark\footnotetext{
The Infrared Telescope Facility,
is operated by the University of Hawaii under Cooperative Agreement
no. NCC 5-538 with the National Aeronautics and Space Administration,
Science Mission Directorate, Planetary Astronomy Program.} (IRTF),
Submillimeter Array\footnotemark\footnotetext{
The Submillimeter Array is a joint project between the Smithsonian
Astrophysical Observatory and the Academia Sinica Institute of Astronomy
and Astrophysics and is funded by the Smithsonian Institution and the
Academia Sinica.} (SMA),
and James Clerk Maxwell Telescope\footnotemark\footnotetext{
The James Clerk Maxwell Telescope is operated by The Joint Astronomy Centre
on behalf of the Science and Technology Facilities Council of the United
Kingdom, the Netherlands Organisation for Scientific Research,
and the National Research Council of Canada.} (JCMT)
at the Mauna Kea Observatory.
Our observations, detailed in \S2, are centered on the white box in
Figure~\ref{fig.flamingos}. The data, presented in \S3,
reveal a surprising diversity of evolutionary states at the
cluster center, including the discovery of a Class 0 and starless
core neighboring the known infrared sources. We discuss the
implications of this work for understanding cluster formation in \S4
and conclude in \S5.

\section{Observations}
\begin{figure}[t]
\vskip -1.5in
\includegraphics[width=5in,angle=0]{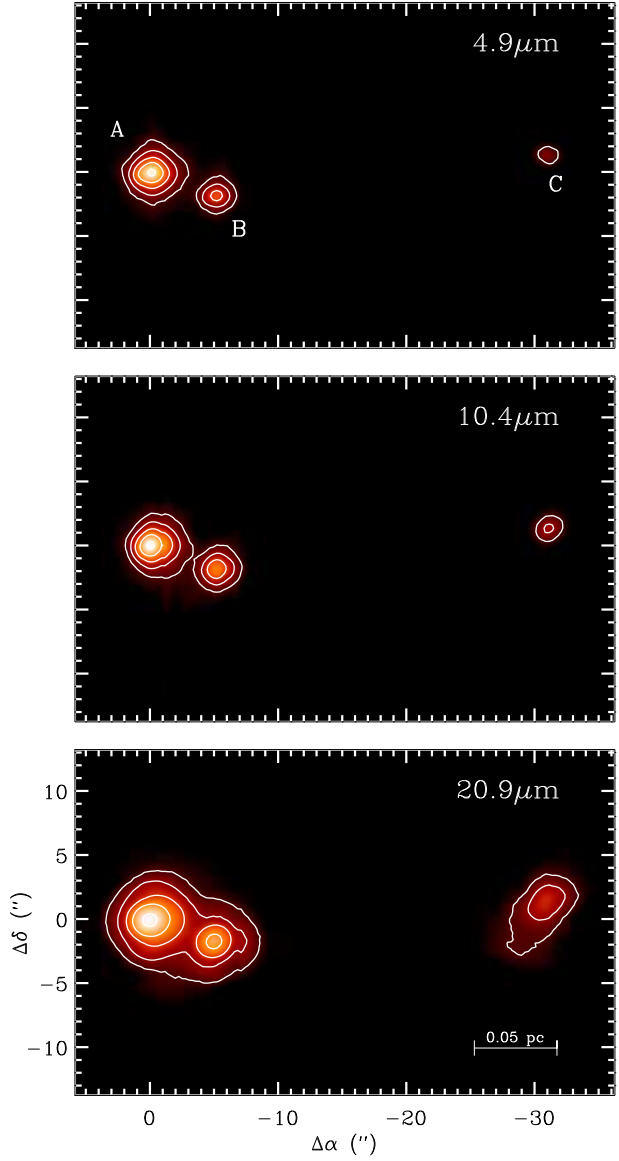}
\caption{MIRSI maps of the center of AFGL961 at M, N, and
Q bands ($4.9, 10.4$, and $20.9\,\micron$ respectively).
For each band, the scale is logarithmic with a dynamic range of 300
and the contour levels are at 1, 3, 9, 27, 81\% of the peak intensity.
The axes are arcsecond offsets from AFGL961A.}
\label{fig.mirsi}
\end{figure}

We used the MIRSI camera on the IRTF \citep{Kassis08}
to map the mid-infrared emission on December 12$^{\rm th}$ 2003.
Images were taken in the M, N, and Q filters centered at
4.9, 10.4, and 20.9\,\micron\ respectively using a $5''\times 5''$ dither.
The skies were clear, cold and dry, and images were
diffraction limited at $0\farcs 5, 0\farcs 9$, and $1\farcs 7$ respectively.
Sky subtraction was obtained by chopping 30'' North-South
and nodding 40'' East-West. This kept the cluster on the array at all
times and maximized the sensitivity in the registered and co-added maps.
Calibration was performed by observations of
$\alpha$ Canis Minor (Procyon) bracketing the AFGL961 observations.
The final maps were made using an IDL reduction pipeline written by
support astronomer E. Volquardsen.
Three sources were detected (Figure~\ref{fig.mirsi})
and all have counterparts in the 2MASS catalog.
We therefore used the latter to define the astrometry.

The SMA observations were carried out under dry and stable skies on
13$^{\rm th}$ December 2005 in the compact configuration ($20-70$\,m
baselines) and in a partial track on 4$^{\rm th}$ February 2006
in the extended configuration ($70-240$\,m baselines).
A mosaic of two Nyquist-spaced pointings was made to ensure
approximately uniform coverage over the MIRSI field.
The receiver was tuned to place the \h2co\ $3_{12}-2_{11}$ line
at 225.7\,GHz in the upper sideband and DCN 3--2 at 217.2\,GHz
in the lower sideband with a resolution of 0.81\,MHz (1.1\,\kms) per channel.
The time dependence of the gains was
measured via observations of 0530+135 interleaved with the source,
the shape of the passband was measured by deep observations of
3C454.3 and 3C111,
and observations of Uranus were used to set the flux scale.
The visibilities were calibrated with the MIR software package
and maps were then produced using standard MIRIAD routines.
A continuum image centered at 1400\,\micron\ was produced
by combining the 2\,GHz lower and upper sidebands from the
compact and extended configuration datasets and inverting
with natural weighting. The resulting resolution and noise is
$3\farcs 1\times 2\farcs 8$ and 3\,mJy\,beam\e.
We also produced a higher resolution map, $1\farcs 4\times 0\farcs 9$,
at the expense of increased noise,
by inverting with higher weights on the longer baselines.
In this paper, we use the former to measure the
distribution and masses of the cores, and the latter
to measure core locations and determine their association
(or lack thereof) with mid-infrared sources.
The lines were undetected on the long baselines and maps were
made using the compact configuration data only using natural weighting.
The resolution and noise in the \h2co\ map are
$3\farcs 7\times 3\farcs 4$ and 0.17\,K.
The corresponding numbers for the DCN map are
$3\farcs 8\times 3\farcs 6$ and 0.08\,K.

As interferometers are unable to measure the flux at small
spatial frequencies, we used Receiver A on the JCMT to observe the same
\h2co\ line as the SMA observations.
We mapped a $3'\times 3'$ region centered on the peak of the SMA map
on 19$^{\rm th}$ March 2008. The map was made using the
on-the-fly mapping mode with a nearby reference
position that was first verified to be free of emission.
Data reduction was carried out with the STARLINK package.
The JCMT map and SMA dirty map were weighted by their respective beam areas,
combined in the image plane, and cleaned with a similarly combined beam
following the method detailed in \cite{Stanimirovic02}.
The resulting map retains the high resolution in the interferometer map
and, as the signal-to-noise ratio was very high in the JCMT data,
the noise is similar, 0.17\,K per channel, as the SMA data.

\section{Results}
\subsection{Morphology}
The MIRSI maps are shown in Figure~\ref{fig.mirsi}.
The three infrared sources first seen by \cite{Lenzen84}
are detected in each filter. To avoid confusion,
particularly for the varied ``western'' nomenclature,
we label them AFGL961A, B and C in order of infrared luminosity.
The positions and fluxes of each source are listed in
Table~\ref{tab.mirsources}.
The fluxes of each source sharply rise with wavelength indicating
that they are deeply embedded. The infrared spectral energy distributions
(SEDs) are discussed in more detail in \S4;
we focus here on the morphological differences.
AFGL961C is slightly elongated in the N-band image and the emission
at Q-band is very extended. \cite{Aspin98} found H$_2$ bow shock features
on either side of the star and \cite{Li08} present a detailed study
of these features. The position angle of the elongated stucture in
the Q-band image lines up with these and the extended mid-infrared
emission is likely to be from hot dust filling a cavity that the
star has blown out around itself.

\begin{figure}[t]
\vskip -1in
\includegraphics[width=6in,angle=0]{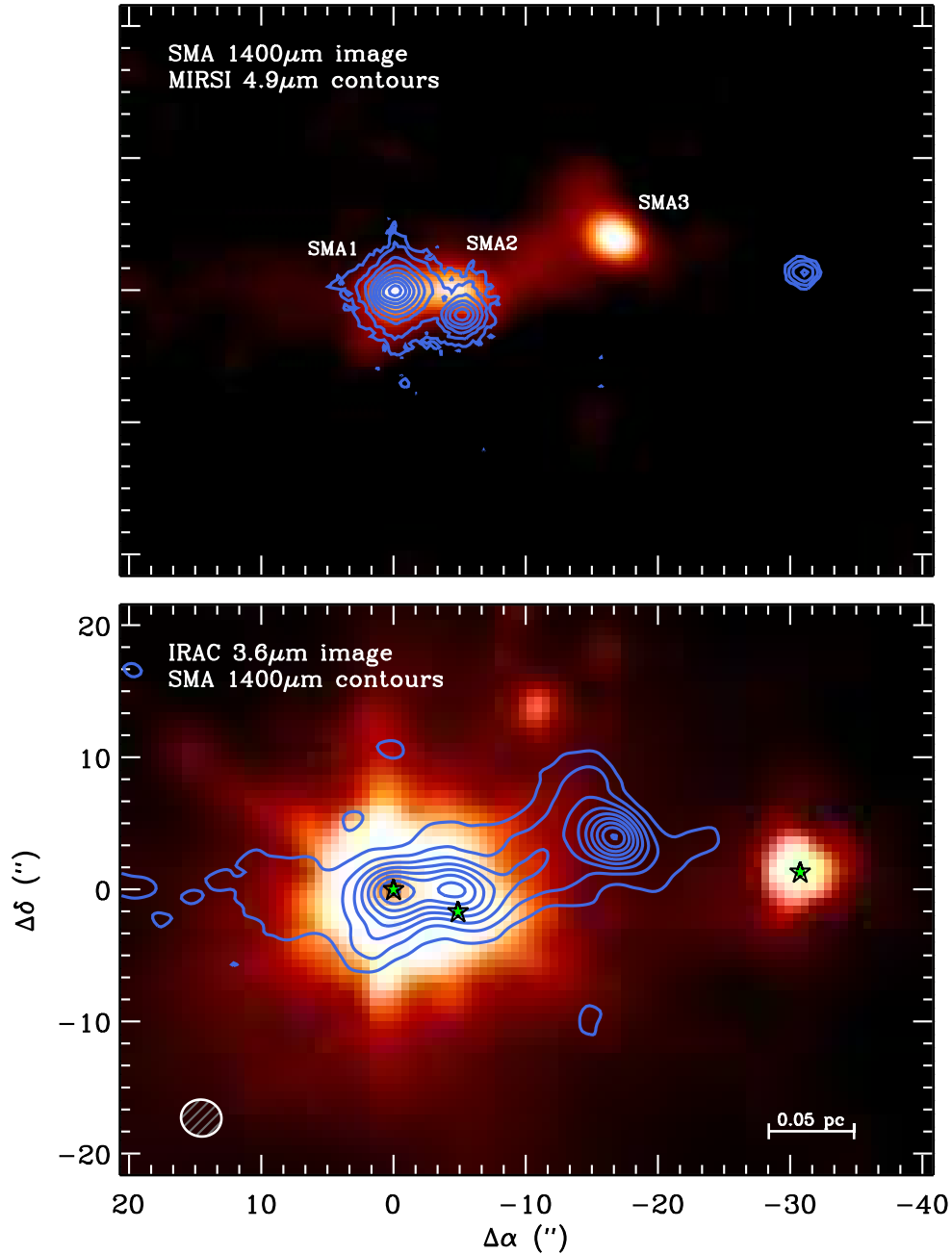}
\vskip -0.5in
\caption{The 1400\,\micron\ continuum emission from the SMA
data showing the cool dust condensations in the cluster.
The top panel overlays the MIRSI M-band image in logarithmic contours
on the 1400\,\micron\ continuum map. The scale ranges
linearly from 5 to 75\,mJy\,beam\e, corresponding to
surface densities $\Sigma=0.05$ to 0.75\,g\,cm\ee.
The three prominent millimeter peaks are labeled SMA1--3 and
their properties listed in Table~\ref{tab.mmsources}.
The bottom panel overlays the 1400\,\micron\ continuum map
in contours on the Spitzer 3.6\,\micron\ map in log scale.
The contours are at surface densities
$\Sigma=0.1\times(1,2,3...)$\,g\,cm\ee.
The stars show the locations of the MIRSI sources AFGL961A,B,C.
The Spitzer image is saturated at the positions of these
sources but provides stringent limits on the luminosity of any
embedded source in SMA3.
The $3\farcs 1\times 2\farcs 8$ SMA beam is shown in the lower left corner.}
\label{fig.sma}
\end{figure}

The differences between the sources are even more striking at 1400\,\micron.
Figure~\ref{fig.sma} shows the SMA continuum map
in relation to the infrared sources.
As for the SCUBA map in Figure~\ref{fig.flamingos}, the contour
units are converted from flux per beam to a mass surface density
assuming $T=20$\,K and a \cite{Hildebrand83} dust opacity,
$\kappa=0.018$\,cm$^2$\,g\e.
Three prominent sources, strung out along a filament,
are detected and labeled SMA1--3.
Their positions and fluxes are listed in Table~\ref{tab.mmsources}.
There is a negligible $0\farcs 2$ offset between the peak of the
SMA1 core and AFGL961A.
The $1\farcs 6$ offset between SMA2 and AFGL961B is significant, however.
A closeup of the region with the higher resolution continuum map
is discussed in \S\ref{sec.dynamics} and more clearly
shows the distinction between these two objects.

We therefore consider SMA1 to be the dusty envelope around
the pre-main-sequence B star AFGL961A and suggest that SMA2 is a
distinct source in the cluster, a dense core that lacks
an infrared counterpart at the limits of detection in the MIRSI map.
SMA3 is the brightest source in the 1400\,\micron\ map and
also lacks an infrared counterpart.
Both these sources are also undetected in 2MASS images and the
deeper JHK observations of \cite{Roman08}.

The MIRSI data show that any embedded object in SMA2 or SMA3
is more than 500 times fainter than AFGL961A from 4.9-20.9\,\micron\
but this does not rule out a solar mass protostar.
For the case of the relatively isolated SMA3 core,
we can place far more stringent limits on the infrared luminosity
from the Spitzer observations of \cite{Poulton08}.
We plot the SMA continuum map in contours over the
IRAC 3.6\,\micron\ image in the lower panel of Figure~\ref{fig.sma}.
No infrared source is apparent at the position of SMA3 in
this image or in the other IRAC bands.
Comparing with faint stars in the cluster and taking into account the
point spread function and extended nebulosity,
we estimate the limits on the flux of any embedded
object in SMA3 to be 0.4, 0.8, 1.2, and 2\,mJy at 3.6, 4.5, 5.8,
and 8.0\,\micron\ respectively.
Many of the low mass, moderately embedded protostars in Perseus
\citep{Jorgensen06} would have been detected at this level.
SMA2 is lost in the glare around AFGL961A,B
of the IRAC images so these limits do not apply in that case.

The Q-band data are critical for ruling out very deeply embedded protostars.
Based on the SED models of \cite{Robitaille06}, we estimate that
the most luminous object consistent with our MIRSI upper limits in
SMA2 and SMA3 is a $300\,L_\odot$ protostar behind $\simgt 100$
magnitudes of visual extinction. Unfortunately the Spitzer MIPS
24\,\micron\ image is heavily saturated and the SMA2 and SMA3 cores
lie in the point spread function wings of AFGL961A,B.
\cite{deWit09} recently imaged the AFGL961A,B pair at 24.5\,\micron\ 
with the 8\,m Subaru telescope. Their map also shows no source within
SMA2 but the field-of-view is too small to include SMA3.
The resolution of these data is higher than our MIRSI Q-band image
but the sensitivity does not appear to be significantly greater.

At the other extreme of the MIRSI-SMA comparison,
the bright infrared source AFGL961C is not
detected in the 1400\,\micron\ map.
This indicates a lack of compact, cool dust around it and,
despite the similarities of the mid-infrared spectral slopes,
places it at a more advanced evolutionary state than AFGL961A
which is fully embedded in a cold molecular core.

The millimeter fluxes can be converted directly to a mass
assuming a temperature and dust grain opacity.
Masses are listed in Table~\ref{tab.mmsources}
under the same prescription for the envelope mass calculation in \S1.
The three cores have similar fluxes and, hence, similar inferred masses.
However, we might expect SMA1, which contains a B star, to be hotter
than SMA2 and 3 and its mass to be proportionately lower than listed here.

The principal finding from our examination of the IRTF and SMA maps
is the detection of 5 distinct pre- or proto-stellar
sources within the central regions of the cluster.
AFGL961\,A is embedded in a dense core, AFGL961\,B and C lack
a similarly sized cold dusty envelope, and there are two moderately
massive cores, SMA2 and 3 that lack bright infrared sources.

\subsection{Dynamics}
\label{sec.dynamics}
\begin{figure}[t]
\vskip -1in
\includegraphics[width=6in,angle=0]{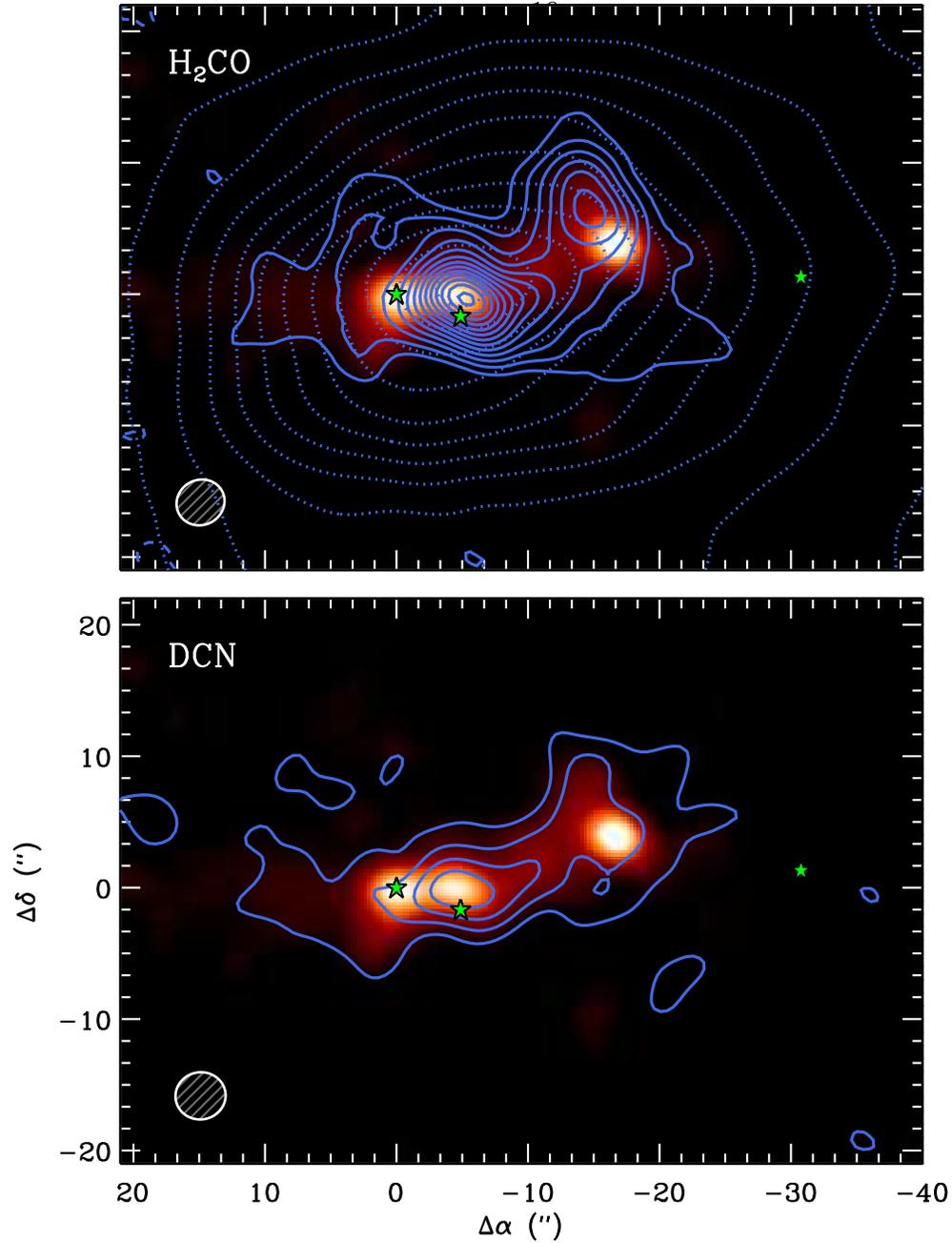}
\vskip -0.5in
\caption{Integrated intensity maps of \h2co\ $3_{12}-2_{11}$
(top panel) and DCN $3-2$ (bottom panel).
The background image in each case is the 1400\,\micron\ continuum map.
The velocity range of integration was 8 to 20\,\kms.
For the \h2co\ map, the SMA data are shown in solid contours
beginning at and in increments of 0.4\,K\,\kms.
Dashed contours show negative levels.
The combined JCMT+SMA map is shown in dotted contours at 4\,K\,\kms.
The DCN contours are at 0.2\,K\,\kms.
The stars shows the location of AFGL961A,B, and C.
The resolution of the integrated intensity maps is shown in
the lower left corner of each panel.}
\label{fig.linemaps}
\end{figure}

Our SMA data provide spectroscopic information at $\sim 1$\,\kms\
resolution and allow us to study the dynamics of the dusty
cores detected in the continuum.
We chose a tuning that placed 3--2 rotational transitions of
\h2co\ in the upper sideband and DCN in the lower sideband.
The former molecule is abundant in protostellar envelopes and
a good tracer of infall motions \citep{Mardones97},
the latter, as a deuterated species, is well suited for pinpointing
the cold, potentially pre-stellar gas \citep{Stark99}
and measuring the systemic motions of the core.
Figure~\ref{fig.linemaps}
shows the integrated intensity of \h2co\ $3_{12}-2_{11}$ and DCN 3--2
overlaid on the continuum. Both lines follow the same filamentary
structure as the dust. The \h2co\ map shows a very strong peak toward
SMA2 and a weaker peak toward SMA3.
The small offsets between the line and dust peaks is likely due to
the high opacity of this line. The DCN emission is weaker but
more closely follows the dust morphology. There is a single peak
toward SMA2 but enhanced emission toward SMA1 and 3 is also evident.

The close correspondence between the DCN and continuum maps shows
that we can use the former to estimate the velocity dispersion,
$\sigma$, of the SMA1--3 cores.
The central velocities, linewidths, $\Delta v=2.355\sigma$,
and virial masses, $M_{\rm vir}=3R\sigma^2/G$,
are listed in Table~\ref{tab.mmsources}.
The latter assumes spherical cores with an inverse square density
profile \citep{Bertoldi92}.
At the resolution of these data, it is hard to determine with
much certainty where the cores end and the filament begins
from the continuum map in Figure~\ref{fig.sma} but, depending
on the intensity threshold used to define their limits, we estimate
core radii $\sim 3-4''$ from their projected area indicating that
the cores are barely resolved with deconvolved sizes $\simlt 5000$\,AU.
We find that the virial masses of SMA1 and 3 exceed the measured
masses by about a factor of two indicating an approximate balance between
kinetic and potential energy in these two cores.
The large virial mass of SMA2, due its high linewidth, indicates
that this core is unbound.

\begin{figure}[t]
\vskip -1in
\includegraphics[width=5in,angle=90]{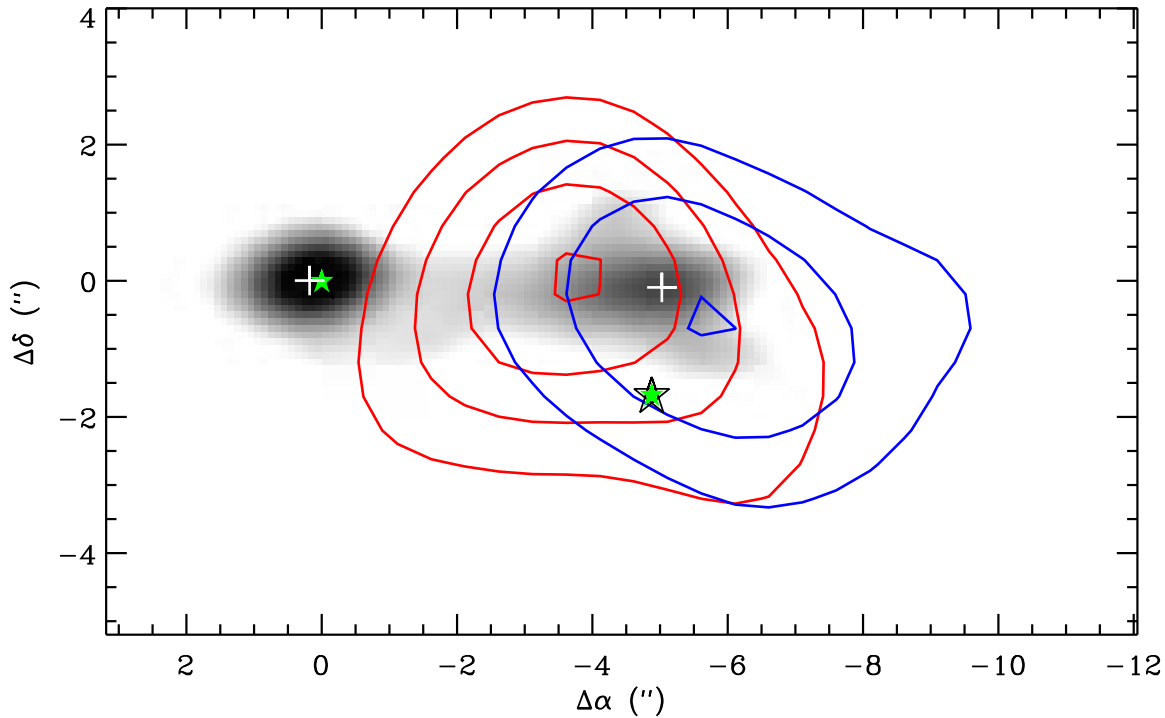}
\caption{Red and blueshifted \h2co\ emission revealing a molecular
outflow from an (undetected) embedded protostar in SMA2.
The background image is the high resolution ($1\farcs 4\times 0\farcs 9$)
1400\,\micron\ continuum map and ranges from 15 to 30\,mJy\,beam\e.
The solid or red contours show the intensity integrated over $13-25$\,\kms.
and the dashed or blue contours show the intensity integrated over $0-13$\,\kms.
Contour levels are 30, 40, 50,... K\,\kms, and show only the
most intense emission above the cloud background.
The stars show the location of AFGL961A, B and the crosses
locate the peak of the millimeter cores SMA1, 2.}
\label{fig.outflow}
\end{figure}

The intense \h2co\ emission toward SMA2 is likely due to grain mantle
sputtering or evaporation and signposts an embedded protostar.
Indeed, inspection of the spectra and channel maps reveal an outflow centered
on this core with line wings discernible from $4-22$\,\kms.
Figure~\ref{fig.outflow} zooms in on this region and shows the
most intense emission from the blue and redshifted sides of the
line. The two moment maps are offset from each other and on opposite
sides of the continuum emission. The discovery of this outflow explains
the high DCN linewidth, why the core is unbound,
and confirms that SMA2 is a physically distinct object from AFGL961B.

\begin{figure}[t]
\vskip -1in
\includegraphics[width=5in,angle=90]{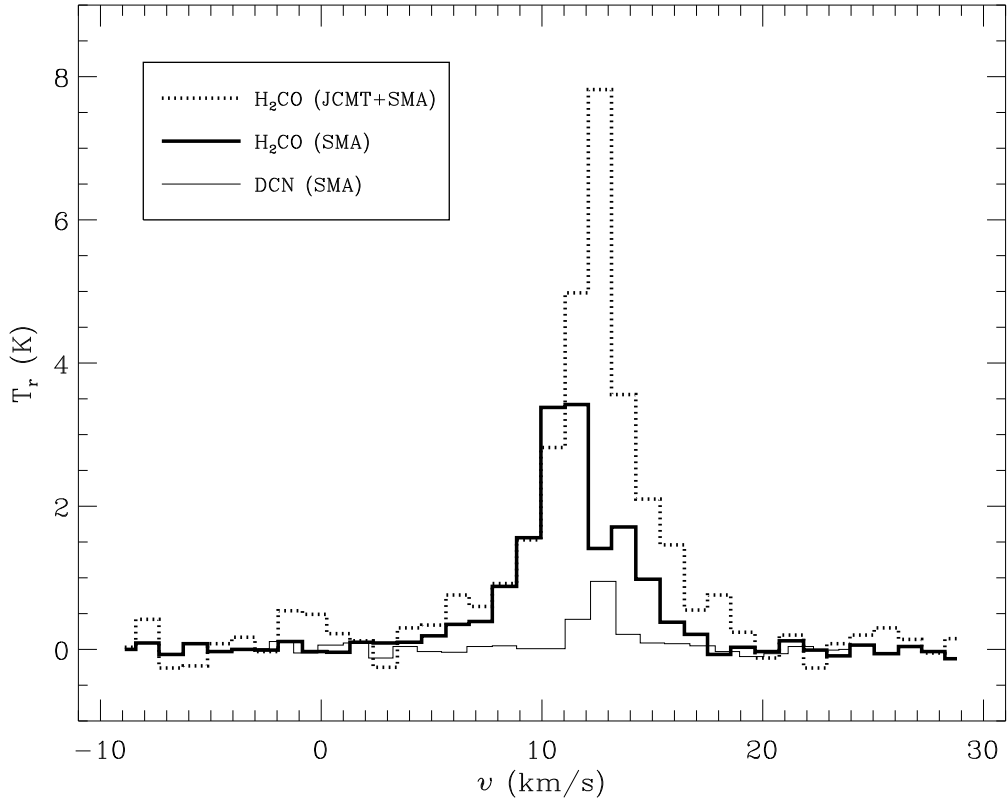}
\caption{Spectra toward SMA3 showing infall at small scales,
possibly due to core collapse.
The thick solid line is the SMA \h2co\ spectrum and shows a slight dip at
the same velocity as the peak of the DCN spectrum shown as a thin solid line.
The dotted line is the combined JCMT+SMA \h2co\ spectrum which
is more symmetric and does not show any self-absorption features.}
\label{fig.spec}
\end{figure}

SMA3 also lacks a detectable infrared source but has a much smaller
linewidth and appears to be bound.
The \h2co\ data does not show any evidence for an outflow but
the spectrum toward the core is asymmetric and shows a small dip at the
velocity of the DCN line (Figure~\ref{fig.spec}).
This is indicative of red-shifted
self-absorption and a signature of infall motion.
Unfortunately the velocity resolution of these data is poor
and the self-absorption is seen in only one spectra channel.
We were consequently unable to successfully fit the spectra
using the radiative transfer models of \cite{DeVries05}.
As there is some ambiguity in modeling the interferometric
data alone without knowledge of the larger scale emission,
we observed the same line with the JCMT and made a fully sampled
\h2co\ datacube. The integrated intensity is shown
as dotted contours in Figure~\ref{fig.linemaps}.
The combined JCMT+SMA map show the large scale cluster envelope
emission, similar to the SCUBA map in Figure~\ref{fig.flamingos}
and the individual pre- and proto-stellar cores are not apparent.
No self-absorption is seen in the combined spectra across the cluster
and toward SMA3 in particular. Hence there is no evidence from these
data for large scale collapse onto the cluster.

\section{Discussion}
\subsection{Evolutionary states}
\begin{figure}[t]
\vskip -2in
\includegraphics[width=4in,angle=0]{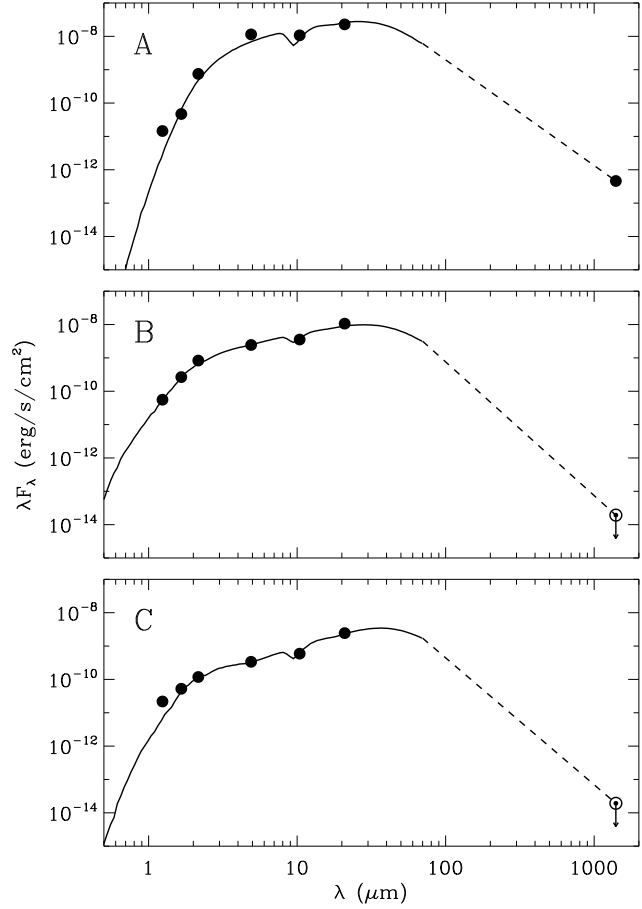}
\caption{Spectral energy distributions
and best fit model for the three infrared sources AFGL961A,B,C.
The dotted line linearly interpolates between the model at 70\,\micron\
and SMA data point at 1400\,\micron.}
\label{fig.sed}
\end{figure}

Our IRTF images from $5-20$\,\micron\ show the most deeply embedded,
luminous stars in the cluster.
The sensitivity of these images is much lower than the Spitzer data
but the resolution is higher and the fidelity of the bright sources better.
We do not find any new sources that are not seen in existing near-infrared
images but we show that the three known members, AFGL961A, B, and C,
each have rising spectral indices in the mid-infrared.
The SEDs are plotted in Figure~\ref{fig.sed}. We have fit the
near- and mid-infrared data using the precomputed models of
\cite{Robitaille06,Robitaille07}. As the model SEDs are noisy at
millimeter wavelengths we have not fit the SMA data point but show the
interpolation between model and 1400\,\micron\ flux.
The model parameters include a central source, disk, and envelope.
In each case, most fits show the sources are embedded under
more than 20\,magnitudes of extinction. The precise amount is not
well constrained (although background sources are ruled out).
Similarly, a wide range of disk and envelope parameters can fit the
data and their masses are not well determined. However, all the fits
show that the envelope dominates, typically by more than a factor of 100.
The best constrained parameters are the total luminosity and
stellar mass of the source. These are tabulated in Table~\ref{tab.sedfits}
for the best fit and the mean and standard deviation of the top
100 fits, weighted by the inverse of $\chi^2$.
The three sources are all massive stars but the
morphological differences and comparison with the millimeter map
shows that they are in quite different evolutionary states.

AFGL961A,B have been considered a binary system based on their close
proximity to each other \citep[e.g.][]{Aspin98}.
However, their properties at 1400\,\micron\ are quite different:
source A is embedded in the dense $6\,M_\odot$ SMA1 core
but there is no millimeter emission centered on source B.
Using the higher resolution continuum map (Figure~\ref{fig.outflow})
we can place a $3\sigma$ limit on the 1400\,\micron\ flux
toward source B of 9\,mJy ($0.3\,M_\odot$) at $\sim 2000$\,AU scales.
AFGL961A,B did not form from a common core, therefore, and are not a
binary pair but simply neighboring protostars, in different
evolutionary states, in a dense cluster.

AFGL961C is different in its own way. It is a point source in the
M-band image, noticeably elongated
at N-band, and a large cavity of hot dust is seen in the Q-band image.
This star is clearing out its surrounding material, as also shown
by the shocked H$_2$ emission image of \cite{Aspin98}.
\cite{Li08} postulated that the hourglass shape of the infrared
nebula is due to polar winds from a very young protostar
punching a hole in its surrounding core.
No core emission is detected at 1400\,\micron, however,
and the same $0.3\,M_\odot$ limit at $\sim 2000$\,AU scales
applies as for source B.

The ratio of stellar to envelope mass, $M_*/M_{\rm env}$,
is an effective measure of protostellar evolutionary state \citep{Andre93}.
There is certainly some uncertainty in determining each of these quantities
but it is clear that there is a large range in the ratio.
$M_*/M_{\rm env}\approx 2$ for AFGL961A but is greater
than 30 for AFGL961B and greater than 15 for AFGL961C.

The outflow toward SMA2 indicates an embedded protostar and the
non-detection in the infrared limit the bolometric luminosity to
less than $300\,L_\odot$. Depending on its age, this constrains
the stellar mass to no more than $6\,M_\odot$
and probably much lower \citep{Siess00}.
Therefore $M_*/M_{\rm env}\simlt 1$ and this is a Class 0 object.
Given its youth and the bipolar \h2co\ structure we see toward it,
we further suggest that this source is the driving source of the
energetic CO outflow from the cluster \citep{Lada82,Dent09}
and not the more luminous infrared sources.

SMA3 is far enough offset from the bright cluster center
for the Spitzer/IRAC data to provide the most stringent limits
on any embedded object and it appears to be truly starless.
As it is gravitationally bound and very dense with
a free-fall time, $t_{\rm ff}\sim 10^4$\,yr,
it is likely to be on the brink of star formation.
The interferometric \h2co\ spectrum shows evidence that the core
is collapsing but higher spectral resolution data are
required to confirm this, and to allow modeling and a
determination of the infall speed. 
The difference between the peak and self-absorption dip is an
approximate measure and suggests $v_{\rm in}\sim 1$\,\kms.
Adding in the short spacing information from the JCMT
swamps the self-absorption showing that any inward motions
are on small scales. That is, the core is collapsing on itself
rather than growing through the accretion of envelope material.
With a mass of $6\,M_\odot$, SMA3 is more massive than starless cores in
isolated star forming regions such as Taurus \citep{Shirley00}.
Presumably it will form a correspondingly more massive star
but probably not comparable to AFGL961A.

A mix of early evolutionary states in clusters is not uncommon.
The first core identified as a Class 0 object,
VLA1623 in $\rho$\,Ophiuchus, sits on the edge of a small
group of pre-stellar cores \citep{Andre93}.
\cite{Williams99} found a collapsing starless core adjacent to
a Class I protostar in the Serpens cluster and
\cite{Swift08} found a range of young stellar objects,
from Class 0 to III, in L1551.
These are all low mass objects in low mass star forming regions.
Radio observations of the W3 and W75\,N massive star forming regions show
a range of HII region morphologies and spectral indices indicating
different evolutionary states \citep[respectively]{Tieftrunk97,Shepherd04}.
In particular, \cite{Shepherd04} were able to estimate an age
spread of at least 1\,Myr between a cluster of five early B stars
based on Str\"omgren sphere expansion.
The detection of cold, dense pre-stellar cores and young lower
mass protostars require observations at millimeter wavelengths.
Interferometer maps by
\cite{Hunter06} and \cite{Rodon08} show tight groups of
dusty cores in NGC6334\,I and W3\,IRS5 respectively.
Core separations are even smaller and masses higher than we
have found here in the lower luminosity AFGL961 cluster.
The cores have a range of infrared properties and some
power outflows and they likely span a range of evolutionary states.
There does not appear, however, to be a clear counterpart to
the pre-stellar core, AFGL961-SMA3, with its combination of
low limit to the luminosity of any embedded source and lack of outflow.

\subsection{Dynamic or equilibrium cluster formation?}
Theoretical models of cluster formation divide into two camps:
a global collapse of a massive molecular clump or piecemeal growth
from the formation of individual protostars in a more local process.
The former occurs on short, dynamical time-scales \citep{Bate03}
but the latter is more gradual and the cluster forming
clump is in quasi-equilibrium \citep{Tan06}.

Based on the SCUBA image in Figure~\ref{fig.flamingos},
the cluster envelope has a mass $M\approx 90\,M_\odot$
within a radius $R\approx 0.25$\,pc.
The free-fall timescale for this region, based on the
inferred average density, $\rho\approx 10^{-19}$\,g\,cm\eee,
is $t_{\rm ff}=(3\pi/32G\rho)^{1/2}\approx 0.2$\,Myr.
Note that material on larger scales, for example in the surrounding
CO clump \citep{Williams95}, would have a longer collapse timescale.

Our observations of a filamentary
structure at high surface densities, $\Sigma\approx 0.1-1$\,g\,cm\ee,
in a roughly spherical envelope at $\Sigma\approx 0.025-0.25$\,g\,cm\ee\
are quite similar to the numerical simulations by
\cite{Bate03} and \cite{Bonnell03}.
In particular, the initial conditions of \cite{Bate03}
with a $M=50\,M_\odot, R=0.19$\,pc clump are closest to the
properties of AFGL961. In these simulations, the clump collapses
on the free-fall timscale,
fragments, and produces stars in two short bursts of
about $0.02$\,Myr duration spread by about 0.2\,Myr.
The addition of radiative feedback \citep{Bate09}
does not change these numbers significantly.

\cite{Kurosawa04} calculated the radiative transfer for the
\cite{Bate03} simulation and showed that the infrared
classification of young stellar objects varied from Class 0 to III.
Under this scenario, a protostar's evolutionary state
is dependent more on its dynamical history than its age
and we would interpret the lack of circumstellar material around
AFGL961B,C as due to their ejection from the dense star forming filament.

\cite{Evans09} shows that circumstellar material around isolated low mass
stars in nearby star forming regions is lost in about 0.5\,Myr.
If the same timescale applies to the more luminous sources in the
crowded environs of AFGL961, then sources B and C are older than the
cluster free-fall time suggesting a quasi-equilibrium collapse.
In this scenario, evolution correlates with age and we would infer
that the massive star AFGL961A formed after B and C, and that collapse
continued even around this compact HII region to form dense cores SMA 2 and 3.

These observations alone cannot distinguish between
dynamic or equilibrium models of cluster formation.
However, similar high resolution mid-infrared through millimeter
observations of other young clusters of varied luminosity
and protostellar density will add to the classification statistics.
In this way, we might hope to decipher the relative effect of
environment and time on protostellar evolution.

\section{Conclusions}
We have carried out a high resolution mid-infrared and millimeter
study of the central region of the young stellar cluster AFGL961.
Our observations show the most deeply embedded bright protostars
in the cluster and the filamentary distribution of the highest
column density gas.
We find five sources within 0.2\,pc of each other,
each with distinct properties unlike the others.
The brightest infrared source is an early B star that powers a compact
HII region and lies within a dense molecular core. The core mass
is a substantial fraction of the stellar mass indicative of a
Class I protostar.
The two other infrared sources are not detected at millimeter wavelengths
and the ratio of circumstellar to stellar mass is very low,
suggesting that these sources are most similar to Class II objects.
Further, one has cleared out a cavity in the circumcluster envelope.
The SMA data also reveal two millimeter cores with no infrared counterparts.
One is a strong source of line emission, drives an outflow,
and has the characteristics of a deeply embedded Class 0 protostar.
The other core is massive, starless, and appears to be collapsing.

The dense mixture of diverse protostellar evolutionary states
suggests either that circumstellar matter is removed rapidly
through dynamical interactions with other cluster members
or that clusters build up gradually over several free-fall timescales.
Regardless of its history, however, the discovery of a massive,
collapsing core shows that AFGL961 continues to grow
even after the birth of a massive star.

\acknowledgments
J.P.W. acknowledges the NSF for support through grants AST-0324328
and AST-0607710 and BHO for his support of the NSF.
We thank Matthew Bate and Ian Bonnell for comments.
This publication makes use of data products from the Two Micron All Sky
Survey, which is a joint project of the University of Massachusetts and
the Infrared Processing and Analysis Center/California Institute of
Technology, funded by the National Aeronautics and Space Administration
and the National Science Foundation,
and the facilities of the Canadian Astronomy Data Centre operated by
the National Research Council of Canada with the support of the
Canadian Space Agency.

\clearpage
\begin{deluxetable}{lcccccccc}
\tablecolumns{9}
\tablewidth{0pc} 
\tablecaption{Infrared sources\label{tab.mirsources}}
\tablehead{
\colhead{} & \colhead{$\Delta\alpha$\tablenotemark{a}} & \colhead{$\Delta\delta$\tablenotemark{a}} & \colhead{$J$} & \colhead{$H$} & \colhead{$K$} & \colhead{$M$} & \colhead{$N$} & \colhead{$Q$} \\
\colhead{Source} & \colhead{($''$)} & \colhead{($''$)} & \colhead{(Jy)} & \colhead{(Jy)} & \colhead{(Jy)} & \colhead{(Jy)} & \colhead{(Jy)} & \colhead{(Jy)}}
\startdata
AFGL961A &    0.0  &   0.0  & 0.006 & 0.023 & 0.009 &  18.7   &  37.2   &  159    \\
AFGL961B &  $-4.9$ & $-1.7$ & 0.026 & 0.147 & 0.029 &  3.98  &  12.3   &   74.0  \\
AFGL961C & $-30.8$ &   1.3  & 0.539 & 0.598 & 0.085 &  0.55  &   2.05  &   17.0  \\
\enddata
\tablenotetext{a}{$\alpha_{2000}=06^{\rm h}34^{\rm m}37.74^{\rm s}, \delta_{2000}=04^\circ\,12'\,44.2''$}
\end{deluxetable}

\begin{deluxetable}{lccccccc}
\tablecolumns{8}
\tablewidth{0pc} 
\tablecaption{Millimeter cores\label{tab.mmsources}}
\tablehead{
\colhead{Source} & \colhead{$\Delta\alpha$} & \colhead{$\Delta\delta$} & \colhead{$S_{\rm 1400}$} & \colhead{$M_{\rm core}$\tablenotemark{a}} & \colhead{$v_{\rm lsr}$} & \colhead{$\Delta v$} & \colhead{$M_{\rm vir}$\tablenotemark{b}} \\
\colhead{} & \colhead{($''$)} & \colhead{($''$)} & \colhead{(mJy)} & \colhead{($M_\odot$)} & \colhead{(\kms)} & \colhead{(\kms)} & \colhead{($M_\odot$)}}
\startdata
SMA1   &    0.2  &   0.0  & 215  &   6.3   & 15.0 &  2.1  & 13    \\
SMA2   &  $-5.0$ & $-0.1$ & 184  &   5.4   & 14.2 &  4.3  & 56    \\
SMA3   & $-16.8$ &    4.0 & 210  &   6.2   & 12.6 &  1.8  &  9.8  \\
\enddata
\tablenotetext{a}{$T=20$\,K, $\kappa_\nu=0.19$\,cm$^2$\,g$^{-1}$}
\tablenotetext{b}{$R=5000$\,AU}
\end{deluxetable}

\begin{deluxetable}{lcccc}
\tablecolumns{5}
\tablewidth{0pc} 
\tablecaption{SED fits\label{tab.sedfits}}
\tablehead{
\colhead{} & \multicolumn{2}{c}{$M_\ast~(M_\odot)$} &\multicolumn{2}{c}{$L~(L_\odot)$} \\
\colhead{Source} & \colhead{Best} & \colhead{Mean} & \colhead{Best} & \colhead{Mean}}
\startdata
AFGL961A & 11.0 & $11.3 \pm 1.8$ & 4600 & $6000 \pm 2500$ \\
AFGL961B &  9.1 & $ 9.1 \pm 1.3$ & 1200 & $2500 \pm 1300$ \\
AFGL961C &  5.3 & $ 6.4 \pm 1.1$ &  370 & $ 480 \pm  200$ \\
\enddata
\end{deluxetable}

\clearpage

\clearpage

\clearpage

\clearpage

\clearpage

\clearpage

\clearpage


\begin{thebibliography}{}
\bibitem[Andre et al.(1993)]{Andre93} Andre, P., Ward-Thompson, D., \& Barsony, M.\ 1993, \apj, 406, 122 
\bibitem[Aspin(1998)]{Aspin98} Aspin, C.\ 1998, \aap, 335, 1040 
\bibitem[Bally \& Predmore(1983)]{Bally83} Bally, J., \& Predmore, R.\ 1983, \apj, 265, 778 
\bibitem[Bate et al.(2003)]{Bate03} Bate, M.~R., Bonnell, I.~A., \& Bromm, V.\ 2003, \mnras, 339, 577 
\bibitem[Bate(2009)]{Bate09} Bate, M.~R.\ 2009, \mnras, 392, 1363
\bibitem[Bertoldi \& McKee(1992)]{Bertoldi92} Bertoldi, F., \& McKee, C.~F.\ 1992, \apj, 395, 140
\bibitem[Blitz \& Thaddeus(1980)]{Blitz80} Blitz, L., \& Thaddeus, P.\ 1980, \apj, 241, 676 
\bibitem[Bonnell et al.(2003)]{Bonnell03} Bonnell, I.~A., Bate, M.~R., \& Vine, S.~G.\ 2003, \mnras, 343, 413
\bibitem[Castelaz et al.(1985)]{Castelaz85} Castelaz, M.~W., Grasdalen, G.~L., Hackwell, J.~A., Capps, R.~W., \& Thompson, D.\ 1985, \aj, 90, 1113 
\bibitem[Cohen(1973)]{Cohen73} Cohen, M.\ 1973, \apjl, 185, L75 
\bibitem[Cox et al.(1990)]{Cox90} Cox, P., Deharveng, L., \& Leene, A.\ 1990, \aap, 230, 181 
\bibitem[Dent et al.(2009)]{Dent09} Dent, W.~R.~F., et al.\ 2009, arXiv:0902.4138 
\bibitem[De Vries \& Myers(2005)]{DeVries05} De Vries, C.~H., \& Myers, P.~C.\ 2005, \apj, 620, 800 
\bibitem[de Wit et al.(2009)]{deWit09} de Wit, W.~J., et al.\ 2009, \aap, 494, 157
\bibitem[Evans et al.(2009)]{Evans09} Evans, N.~J., et al.\ 2009, \apjs, 181, 321
\bibitem[Hildebrand(1983)]{Hildebrand83} Hildebrand, R.~H.\ 1983, \qjras, 24, 267 
\bibitem[Hensberge et al.(2000)]{Hensberge00} Hensberge, H., Pavlovski, K., \& Verschueren, W.\ 2000, \aap, 358, 553 
\bibitem[Hunter et al.(2006)]{Hunter06} Hunter, T.~R., Brogan, C.~L., Megeath, S.~T., Menten, K.~M., Beuther, H., \& Thorwirth, S.\ 2006, \apj, 649, 888
\bibitem[J{\o}rgensen et al.(2006)]{Jorgensen06} J{\o}rgensen, J.~K., et al.\ 2006, \apj, 645, 1246 
\bibitem[Kassis et al.(2008)]{Kassis08} Kassis, M., Adams, J.~D., Hora, J.~L., Deutsch, L.~K., \& Tollestrup, E.~V.\ 2008, \pasp, 120, 1271 
\bibitem[Kurosawa et al.(2004)]{Kurosawa04} Kurosawa, R., Harries, T.~J., Bate, M.~R., \& Symington, N.~H.\ 2004, \mnras, 351, 1134 
\bibitem[Lada \& Gautier(1982)]{Lada82} Lada, C.~J., \& Gautier, T.~N., III 1982, \apj, 261, 161 
\bibitem[Lenzen et al.(1984)]{Lenzen84} Lenzen, R., Hodapp, K.-W., \& Reddmann, T.\ 1984, \aap, 137, 365
\bibitem[Li et al.(2008)]{Li08} Li, J.~Z., Smith, M.~D., Gredel, R., Davis, C.~J., \& Rector, T.~A.\ 2008, \apjl, 679, L101 
\bibitem[Mardones et al.(1997)]{Mardones97} Mardones, D., Myers, P.~C., Tafalla, M., Wilner, D.~J., Bachiller, R., \& Garay, G.\ 1997, \apj, 489, 719 
\bibitem[Ogura \& Ishida(1981)]{Ogura81} Ogura, K., \& Ishida, K.\ 1981, \pasj, 33, 149 
\bibitem[Perez et al.(1987)]{Perez87} Perez, M.~R., The, P.~S., \& Westerlund, B.~E.\ 1987, \pasp, 99, 1050 
\bibitem[Poulton et al.(2008)]{Poulton08} Poulton, C.~J., Robitaille, T.~P., Greaves, J.~S., Bonnell, I.~A., Williams, J.~P., \& Heyer, M.~H.\ 2008, \mnras, 384, 1249 
\bibitem[Robitaille et al.(2006)]{Robitaille06} Robitaille, T.~P., Whitney, B.~A., Indebetouw, R., Wood, K., \& Denzmore, P.\ 2006, \apjs, 167, 256
\bibitem[Robitaille et al.(2007)]{Robitaille07} Robitaille, T.~P., Whitney, B.~A., Indebetouw, R., \& Wood, K.\ 2007, \apjs, 169, 328
\bibitem[Rod{\'o}n et al.(2008)]{Rodon08} Rod{\'o}n, J.~A., Beuther, H., Megeath, S.~T., \& van der Tak, F.~F.~S.\ 2008, \aap, 490, 213
\bibitem[Rom{\'a}n-Z{\'u}{\~n}iga et al.(2008)]{Roman08} Rom{\'a}n-Z{\'u}{\~n}iga, C.~G., Elston, R., Ferreira, B., \& Lada, E.~A.\ 2008, \apj, 672, 861 
\bibitem[Shepherd et al.(2004)]{Shepherd04} Shepherd, D.~S., Kurtz, S.~E., \& Testi, L.\ 2004, \apj, 601, 952 
\bibitem[Shirley et al.(2000)]{Shirley00} Shirley, Y.~L., Evans, N.~J., II, Rawlings, J.~M.~C., \& Gregersen, E.~M.\ 2000, \apjs, 131, 249 
\bibitem[Siess et al.(2000)]{Siess00} Siess, L., Dufour, E., \& Forestini, M.\ 2000, \aap, 358, 593
\bibitem[Stark et al.(1999)]{Stark99} Stark, R., van der Tak, F.~F.~S., \& van Dishoeck, E.~F.\ 1999, \apjl, 521, L67 
\bibitem[Stanimirovic(2002)]{Stanimirovic02} Stanimirovic, S.\ 2002, Single-Dish Radio Astronomy: Techniques and Applications, 278, 375
\bibitem[Swift \& Welch(2008)]{Swift08} Swift, J.~J., \& Welch, W.~J.\ 2008, \apjs, 174, 202
\bibitem[Tan et al.(2006)]{Tan06} Tan, J.~C., Krumholz, M.~R., \& McKee, C.~F.\ 2006, \apjl, 641, L121
\bibitem[Tieftrunk et al.(1997)]{Tieftrunk97} Tieftrunk, A.~R., Gaume, R.~A., Claussen, M.~J., Wilson, T.~L., \& Johnston, K.~J.\ 1997, \aap, 318, 931 
\bibitem[Williams et al.(1995)]{Williams95} Williams, J.~P., Blitz, L., \& Stark, A.~A.\ 1995, \apj, 451, 252
\bibitem[Williams \& Myers(1999)]{Williams99} Williams, J.~P., \& Myers, P.~C.\ 1999, \apjl, 518, L37 
\bibitem[Zinnecker \& Yorke(2007)]{Zinnecker07} Zinnecker, H., \& Yorke, H.~W.\ 2007, \araa, 45, 481 
\end{thebibliography}
\end{document}